\documentclass[twocolumn,aps,showpacs,amsmath,amssymb, superscriptaddress]{revtex4-1}

\usepackage{graphicx}
\usepackage{bm}
\usepackage{pstricks,pst-node,pst-text,pst-3d,pst-plot}

\newcommand{\la}{\langle}
\newcommand{\ra}{\rangle}
\begin{document}

\title{Accurate expansions of internal energy and specific heat of critical
two-dimensional Ising model with free boundaries}
\author{Xintian Wu}
\email{wuxt@bnu.edu.cn}
\affiliation{Department of Physics, Beijing Normal University,
Beijing, 100875, China}

\author{Ru Zheng}
\affiliation{Department of Physics, Beijing Normal University,
Beijing, 100875, China}

\author{Nickolay Izmailian}
\email{izmail@yerphi.am}
\affiliation{Applied Mathematics Research Centre, Coventry University, Coventry CV1
5FB, United Kingdom} \affiliation{A.I. Alikhanyan National
Science Laboratory, Alikhanian Br.2, 375036 Yerevan, Armenia.  }
\author{ Wenan  Guo }
\email{waguo@bnu.edu.cn}
\affiliation{Department of Physics, Beijing Normal University,
Beijing, 100875, China}

\date{\today}

\begin{abstract}

The bond-propagation (BP) algorithm for the specific heat of the
two dimensional Ising model is developed and that for the internal
energy is completed. Using these algorithms, we study the critical
internal energy and specific heat of the model on the square
lattice and triangular lattice with free boundaries. Comparing
with previous works [X.-T. Wu {\it et al} Phys. Rev. E {\bf 86},
041149 (2012) and Phys. Rev. E {\bf 87}, 022124 (2013)], we reach
much higher accuracy ($10^{-26}$) of
the internal energy and specific heat,
compared to the accuracy $10^{-11}$ of the internal energy and $10^{-9}$
of the specific heat reached in the previous works.
This leads to much more accurate estimations of 
the edge and corner terms.   
The exact values of some edge and corner terms are therefore
conjectured. The accurate forms of finite-size scaling for the
internal energy and specific heat are determined for the
rectangle-shaped square lattice with various aspect ratios and
various shaped triangular lattice. For the rectangle-shaped square
and triangular lattices and the triangle-shaped triangular
lattice, there is no logarithmic correction terms of order higher
than $1/S$, with $S$ the area of the system. For the triangular
lattice in rhombus, trapezoid and hexagonal shapes, there exist
logarithmic correction terms of order higher than $1/S$ for the
internal energy, and logarithmic correction terms of all orders
for the specific heat.
\end{abstract}


\maketitle

\section{Introduction}

The finite-size scaling theory, introduced by Fisher, finds
extensive applications in the analysis of experimental, Monte
Carlo, and transfer-matrix data, as well as in recent theoretical
developments related to conformal invariance
\cite{privman,privman1,blote,cardy}. Exact solutions on the Ising
model have been used to determine the form of finite-size scaling.
Exact results of the model on finite sizes with various
boundaries have been studied intensively 
\cite{onsager,kaufman,fisher1969,fisher,izmailian2002a,izmailian2002b,izmailian2007,salas,Janke}.
Detailed knowledge has been obtained for the torus case
\cite{izmailian2002a,salas}, for helical boundary conditions
\cite{izmailian2007}, for Brascamp-Kunz boundary conditions
\cite{izmailian2002b,Janke} and for infinitely long cylinder\cite{izmailian1}.

It is well known that the two-dimensional (2D) Ising model has
been studied very intensively. However the solution for a 2D
system with free boundaries, i.e., with free edges and sharp
corners, is difficult and not well studied up to now.  Although
there are Monte Carlo and transfer matrix studies on this problem
\cite{landau, stosic}, the accuracy, or the system sizes achieved,
is not enough to extract the finite-size corrections. Meanwhile,
for 2D critical systems, a huge amount of knowledge has been
obtained by the application of the powerful techniques of
integrability and conformal field theory (CFT) \cite{blote,cardy,
kleban}. Cardy and Peschel predicted that the next sub-dominant
contribution to the free energy on a square comes from the corners
\cite{cardy}, which is universal, and related to the central
charge $c$ in the continuum limit. Along this direction, boundary
conformal field theory has been studied intensively to treat
critical systems with free boundaries in recent years
\cite{bondesan, imamura}. It plays a fundamental role in our
understanding of logarithmic conformal field theory
\cite{Gaberdiel,Read}, of the Kondo effect \cite{Affleck}, of the
physics of quantum impurities or the Fermi edge singularity
\cite{Affleck1}, of local and global quenches in one-dimensional
quantum systems \cite{calabrese, dubail}, and, in the relationship
between conformal field theory and the Schramm Loewner Evolution
formalism \cite{Bauer}. However, till now there is few studies on
lattice model, say Ising model, with free boundary condition to
compare with the CFT results.

Recently  there appear two successful approaches to solve the 2D
Ising model with free boundaries.
Vernier and Jacobsen \cite{jacobsen} conjectured an exact analytic formula for the
corner free energy of the Ising model on the square lattice. The
asymptotic behavior upon approaching the critical temperature
is shown to be consistent with CFT results.
The other way is to use the  bond propagation (BP) algorithm, which was
developed for computing the partition function of the Ising model
in two dimensions  \cite{loh1,loh2}. It is much faster than Monte
Carlo simulation, and costs moderate memory comparing with
the transfer matrix method. Very large system size can be reached.
With this algorithm, the calculations have been carried out on
square and triangular lattices with free boundaries \cite{wu,wu2}. The
results of free energy are surprisingly accurate to $10^{-26}$.
Fitting the assumed finite-size scaling formula to the data, the
edge and corner terms were obtained very accurately. For example, from the
corner term on the square lattice with rectangle shape \cite{wu},  the
central charge of Ising model was estimated as $c=0.5\pm 1\times 10^{-10}$,
compared with the CFT result $c=0.5$ \cite{cardy}.

In the previous work \cite{wu, wu2},
the critical internal energy and specific heat have also been studied
by numerical differentiation of free energy \cite{wu,wu2}.
However, the numerical differentiation process limits the accuracy to
$10^{-11}$ and $10^{-9}$, respectively, which is much lower than that of
the free energy. Therefore the authors could
not determine the accurate expansion form for the internal energy
and specific heat. For example, the authors could not extract the logarithmic
correction at the third and fourth order in the expansions.

In this paper we use the BP algorithm for internal energy and
specific heat, rather than differentiation method, to study the
critical internal energy and specific heat on the square and
triangular lattices with free boundaries. The $Y-\Delta$
transformation in BP algorithm for internal energy was given by
Loh et.al \cite{loh2}. In this paper, we derive the BP series
reduction for the internal energy, and make the algorithm
completed.  We also develop the BP algorithm for specific heat.
Using these algorithms, the accuracy of internal energy and
specific heat reaches the same level as that of free energy, which
is $10^{-26}$. Fitting to these results, we obtain very accurate
expansion of the internal energy and specific heat. The accurate
expansions is helpful to the CFT and renormalization group (RG)
study on this kind of problem \cite{vicari}.

For the square lattice, the rectangular
shape with size $M\times N$ are studied. Five aspect ratios
$\rho=M/N=1,2,4,8,16$ are investigated. It is shown that the edge
and corner internal energy are independent of aspect ratio, and
there is no logarithmic correction term of order higher than
$1/N^2$.  For the triangular lattice, five shapes: triangular,
rhomboid, trapezoid, hexagonal and rectangular, are studied.   The
logarithmic corrections terms $\ln N/N^4,\ln N/N^5,\cdots$ are
found in the internal energy density, and logarithmic correction
terms $\ln N/N^3,\ln N/N^4,\cdots$ are found in the specific heat
for the rhomboid, trapezoid and hexagonal shape, while there are
no such terms for the triangular and rectangular shape. The
exact values of some edge and corner terms are conjectured
with the help of high accurate data.

Our paper is organized as follows. First we derive the BP
algorithm for the specific heat section II. In section III, we
give the result of internal energy and specific heat for the square
lattice in rectangular shape. In section IV, we present results
for the triangular lattice in five shapes. We summarize in Section V.

\section{BP algorithm for specific heat}

The partition function of the Ising model on a 2D lattice is
\begin{equation}
Z=\sum_{\{\sigma\}}e^{-\beta H} \label{eq:partition},
\end{equation}
where the Hamiltonian is given by
\begin{equation}
\beta H=-\sum_{\langle ij \rangle} J_{ij}\sigma_i\sigma_j,
\end{equation}
where $J_{ij}=\beta K_{ij}$ and $K_{ij}=K$ are the homogeneous couplings.
The free energy density is defined as
\begin{equation}
f=\frac{F}{S}=\frac{\ln Z}{S},
\end{equation}
where  $F,S$ are the total free energy and number of spins or the
area of the system respectively.

The internal energy and specific heat density can be computed from
\begin{equation}
u=\frac{U}{S}=\frac{\partial f}{\partial
\beta}=-\frac{1}{SZ}\sum_{\{\sigma\}}(He^{-\beta H})
\end{equation}
and
\begin{equation}
c=\frac{C}{S}=\beta^2 \frac{\partial^2 f}{\partial
\beta^2}=\frac{\beta^2}{S}(\sum_{\{\sigma\}}H^2e^{-\beta
H}/Z-U^2 )\label{eq:specific}
\end{equation}
where $U,C$ are the total internal energy and heat capacity.
Note here
the definition of internal energy is different from the ordinary one
in the sign.

Along the the same route as the BP algorithm for the internal
energy in reference \cite{loh2}, we develop the BP
algorithm for the specific heat.  Using these algorithm, we can avoid numerical
differentiation used in previous works \cite{wu,wu2} and
obtain much accurate results.
For the convenience of the interested readers, we give the
details of our algorithm here.

\begin{figure}
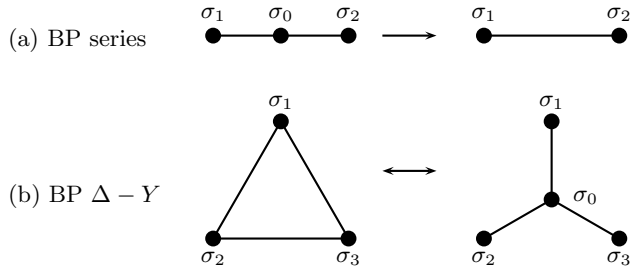

\centering
\psset{xunit=0.9cm}
\psset{yunit=0.9cm}
\pspicture(0,0)(8,4)
\uput[270](0,3.3){(a) BP series}
\uput[270](2,3.6){$\sigma_1$}
\uput[270](3,3.6){$\sigma_0$}
\uput[270](4,3.6){$\sigma_2$}
\psline{<->}(4.5,1)(5.3,1)
\psline{->}(4.5,3.)(5.3,3)
\psline(2,3)(4,3)
\rput{0}(2,3){\pscircle*{3pt}}
\rput{0}(3,3){\pscircle*{3pt}}
\rput{0}(4,3){\pscircle*{3pt}}
\psline(6,3)(8,3)
\uput[270](6,3.6){$\sigma_1$}
\uput[270](8,3.6){$\sigma_2$}
\rput{0}(6,3){\pscircle*{3pt}}
\rput{0}(8,3){\pscircle*{3pt}}
\uput[270](0.1,1){(b) BP $\Delta-Y$}
\uput[270](2,0.0){$\sigma_2$}
\uput[270](4,0.0){$\sigma_3$}
\uput[270](3,2.3){$\sigma_1$}
\rput{0}(2,0){\pscircle*{3pt}}
\rput{0}(3,1.732){\pscircle*{3pt}}
\rput{0}(4,0){\pscircle*{3pt}}
\psline(2,0)(3,1.732)
\psline(3,1.7322)(4,0)
\psline(2,0)(4,0)

\rput{0}(6,0){\pscircle*{3pt}}
\rput{0}(7,1.732){\pscircle*{3pt}}
\rput{0}(8,0){\pscircle*{3pt}}
\rput{0}(7,0.577){\pscircle*{3pt}}
\psline(6,0)(7,0.577)
\psline(7,0.577)(8,0)
\psline(7,0.577)(7,1.732)
\uput[270](6,0.0){$\sigma_2$}
\uput[270](8,0.0){$\sigma_3$}
\uput[270](7,2.3){$\sigma_1$}
\uput[270](7.5,0.9){$\sigma_0$}
\endpspicture
\caption{(Color online) (a) and (b) are building blocks of the BP
algorithm: BP series and BP $\Delta -Y$ transformation,
respectively.
  }
\label{bp}
\end{figure}

The scheme of the BP algorithm is given in \cite{loh1,loh2}.
Two main transformations: BP series reduction and BP
Y-$\Delta$ transformation, as shown in Fig. 1, are
the building blocks. To calculate the free energy the transformations
preserve $Z$. To calculate the internal energy other transformations
preserving $Z'\equiv \Sigma_{\{\sigma\}} H {\rm e}^{-\beta H}$ are needed.
In order to use the BP algorithm to calculate the specific heat
directly, we further need corresponding transformations which preserve
$Z''\equiv \sum_{\{\sigma\}} H^2 {\rm e}^{-\beta H}$.
The procedure is equivalent to do transformations
preserving the following quantity
\begin{equation}
\sum_{\{\sigma\}}[(\sum_{\la ij \ra}J'_{ij}\sigma_i\sigma_j+U)^2
+\sum_{\la ij \ra}J''_{ij}\sigma_i\sigma_j+C]e^{F+\sum_{\la ij \ra}J_{ij}\sigma_i\sigma_j},
\end{equation}
in which new couplings $J'$ and $J''$ are introduced with
initial values $J'_{ij}=K_{ij}, J''_{ij}=0$. $F, U$ and $C$ is
set to zero initially. All couplings and $F, U, C$ change values in each step.
As the transformations are completed, i.e., all bonds are eliminated by the BP
algorithm, the final results of $F, U$ and $C$ are obtained as the
total free energy, internal energy and heat capacity, respectively.

At first, let us discuss the BP ``series" reduction for the specific
heat. It corresponds to integrating out a spin with two neighbors,
generating effective couplings $J_{12},J'_{12},J''_{12}$ such that
\begin{widetext}
\begin{eqnarray}
&  & \sum_{\sigma_0}
e^{J_{10}\sigma_0\sigma_1+J_{20}\sigma_0\sigma_2+H_{\rm ex}}[(J'_{10}\sigma_0\sigma_1+J'_{20}\sigma_0\sigma_2+H'_{\rm ex})^2
 +J''_{10}\sigma_0\sigma_1+J''_{20}\sigma_0\sigma_2+H''_{\rm ex}]
\nonumber \\
 & =&e^{J_{12}\sigma_1\sigma_2+H_{\rm ex}+\delta F} [(J'_{12}\sigma_1\sigma_2+H'_{\rm ex}+\delta
 U)^2   +J''_{12}\sigma_1\sigma_2+H''_{\rm ex}+\delta C],
\label{eq:series}
\end{eqnarray}
\end{widetext}
where the quadratic terms not involving $\sigma_0$ are collected
in $H_{\rm ex}(\sigma_1, \sigma_2, \cdots, F)$, $H'_{\rm ex}(\sigma_1, \sigma_2, \cdots, U)$ and
$H''_{\rm ex}(\sigma_1, \sigma_2, \cdots, C)$, respectively, according to
the couplings.
In each step, we have $F\rightarrow F+\delta F$, $U\rightarrow
U+\delta U$ and $C\rightarrow C+\delta C$. Since
$H'_{\rm ex}$ and $H''_{\rm ex}$ contain the variables
$\sigma_1, \sigma_2, \cdots, U$ and $C$, the prefactors of them in
both sides of the above equation must be equal. Equating the
prefactors of $H'^2_{\rm ex}$, or $H''_{\rm ex}$, in both sides of the
above equation yields
\begin{equation}
\sum_{\sigma_0}e^{J_{10}\sigma_1\sigma_0+J_{20}\sigma_2\sigma_0}=e^{\delta
F+J_{12}\sigma_1\sigma_2}.
\end{equation}
This is the BP series transformation for the partition function
\cite{loh1,loh2},  which requires
\begin{equation}
e^{\delta F\pm J_{12}} =  2 \cosh(J_{10}\pm J_{20}) \equiv a_{\pm}.
\label{eq:a}
\end{equation}
The solution is given by
\begin{equation}
\delta f=\sqrt{a_{+}a_{-}}, \hskip 0.5cm j_{12}=\sqrt{a_{+}/a_{-}},
\label{eq:bp-f}
\end{equation}
with $\delta f=e^{\delta F}$ and $j_{ij}=e^{-J_{ij}}$.

Equating the prefactors of $H'_{\rm ex}$ in the two sides of Eq.~(\ref{eq:series}) yields
\begin{eqnarray}
&&\sum_{\sigma_0}[J'_{10}\sigma_1\sigma_0+J'_{20}\sigma_2\sigma_0]e^{J_{10}\sigma_1\sigma_0+J_{20}\sigma_2\sigma_0}
\nonumber \\
 &=&[J'_{12}\sigma_1\sigma_2+\delta U]e^{\delta
F+J_{12}\sigma_1\sigma_2},
\end{eqnarray}
which leads to
\begin{equation}
e^{\delta F\pm J_{12}}(\delta U \pm J'_{12}) = 2(J'_{10}\pm
J'_{20}) \sinh (J_{10}\pm J_{20}). \label{eq:a'}
\end{equation}
Substituting Eq.~(\ref{eq:a}) into above equation, we obtain
\begin{equation}
\delta U\pm J'_{12} = (J'_{10}\pm J'_{20})\tanh (J_{10}\pm
J_{20})\equiv a'_{\pm},
\end{equation}
with solution given by
\begin{equation}
\delta U=\frac{1}{2}(a'_{+}+a'_{-}), \hskip 0.5cm
J'_{12}=\frac{1}{2}(a'_{+}-a'_{-}). \label{eq:bp-u}
\end{equation}
In fact these transformations are the BP series reduction for the
internal energy. Although the $Y-\Delta$ transformations for the
internal energy has been discussed in Ref. \cite{loh2}, this
transformation was not given.

The rest terms in both sides of Eq. (\ref{eq:series}) also
should be equal, which means
\begin{widetext}
\begin{equation}
\sum_{\sigma_0}e^{J_{10}\sigma_1\sigma_0+J_{20}\sigma_2\sigma_0}
[(J'_{10}\sigma_1\sigma_0+J'_{20}\sigma_2\sigma_0)^2+J''_{10}\sigma_1\sigma_0+J''_{20}\sigma_2\sigma_0]
=e^{\delta
F+J_{12}\sigma_1\sigma_2}[(J'_{12}\sigma_1\sigma_2+\delta
U)^2+J''_{12}\sigma_1\sigma_2+\delta C].
\end{equation}
\end{widetext}
This leads to
\begin{widetext}
\begin{equation}
2(J'_{10}\pm J'_{20})^2\cosh(J_1\pm J_2)+2(J''_{10}\pm
J''_{20})\sinh(J_1\pm J_2)  =e^{\delta F\pm J_{12}}[(\delta U\pm
J'_{12})^2\pm J''_{12}+\delta C].
\end{equation}
\end{widetext}
Substituting Eqs. (\ref{eq:a}) and (\ref{eq:a'}) into the above
equation, we have
\begin{equation}
\delta C = (a''_{+}+a''_{-})/2, ~~~~~
 J''_{12} = (a''_{+}-a''_{-})/2,
 \label{eq:bp-c}
\end{equation}
where
\begin{eqnarray}
a''_{\pm} & = & (J'_{10}\pm J'_{20})^2+(J''_{10}\pm
J''_{20})\tanh(J_{10}\pm J_{20}) \nonumber \\
& &-(a'_{\pm})^{2}.
\end{eqnarray}
Eq. (\ref{eq:bp-f}), (\ref{eq:bp-u}) and (\ref{eq:bp-c}) are the
transformations of the BP series reduction for the specific heat.

The BP Y-$\Delta$ transformation corresponds to integrating the center
spin, generating effective couplings $J_{12}, J_{23}, J_{31}$,
$J'_{12}, J'_{23}, J'_{31}$, and $J''_{12}, J''_{23}, J''_{31}$ such that
\begin{widetext}
\begin{eqnarray}
&&e^{J_{12}\sigma_1\sigma_2+J_{23}\sigma_2\sigma_3+J_{31}\sigma_3\sigma_1+\delta
F}
[(J'_{12}\sigma_1\sigma_2+J'_{23}\sigma_2\sigma_3+J'_{31}\sigma_3\sigma_1+H'_{\rm ex}+\delta
 U)^2 \nonumber \\
 &&+J''_{12}\sigma_1\sigma_2+J''_{23}\sigma_2\sigma_3+J''_{31}\sigma_3\sigma_1+H''_{\rm ex}+\delta
C] \nonumber \\
 &= &\sum_{\sigma_0}
e^{J_{10}\sigma_0\sigma_1+J_{20}\sigma_0\sigma_2+J_{30}\sigma_0\sigma_3}
[(J'_{10}\sigma_0\sigma_1+J'_{20}\sigma_0\sigma_2+J'_{30}\sigma_0\sigma_3+H'_{\rm ex})^2
\nonumber \\
&&+J''_{10}\sigma_0\sigma_1+J''_{20}\sigma_0\sigma_2+J''_{30}\sigma_0\sigma_3+H''_{\rm ex}],
 ~~~~~\forall \sigma_1,\sigma_2,\sigma_3.
\label{eq:cydel}
\end{eqnarray}
\end{widetext}
where
$H'_{\rm ex}(\sigma_1,\sigma_2,\cdots, U)$ and $H''_{\rm ex}(\sigma_1,\sigma_2,\cdots, C)$
collect the quadratic terms not involving $\sigma_0$ with couplings $J'$ and $J''$, respectively. Similarly
the prefactors of $H'_{\rm ex}$ and $H''_{\rm ex}$ in both sides of above
equation should be equal. Equating the prefactor of $H'^2_{\rm ex}$ and
$H''_{\rm ex}$ in both sides of the above equation, we obtain
the solution for $\delta F, J_{12}, J_{23}, J_{31}$
\begin{eqnarray}
\delta f & = & (b_0b_1b_2b_3)^{1/4}, \nonumber \\
j_{12}  & = & [b_1b_2/(b_0b_3)]^{-1/4} ~~~~ (cycl.),
\end{eqnarray}
where $\delta f=e^{\delta F}$, $j_{ij}=e^{-J_{ij}}$, and
\begin{eqnarray}
b_0& = & 2 \cosh(J_{10}+J_{20}+J_{30}),
\nonumber \\
b_1 & = & 2 \cosh(J_{20}+J_{30}-J_{10}) ~~~~(cycl.),
 \label{eq:ydel}
\end{eqnarray}
where ``cylc." means that $j_{31}, j_{12}, b_2, b_3$ are related
to those above by cyclic permutations of the indices 1, 2, 3.

The transformation for $\delta U, j'_{12}, j'_{23}, j'_{31}$ is
given by
\begin{eqnarray}
\delta U & = & \frac{1}{4}(b'_0+b'_1+b'_2+b'_3), \nonumber \\
J'_{12}  & = & \frac{1}{4}(b'_0-b'_1-b'_2+b'_3) ~~~~(cycl.),
\end{eqnarray}
where
\begin{eqnarray}
b'_0 & = & (J'_{10}+J'_{20}+J'_{30})\tanh(J_{10}+J_{20}+J_{30}), \nonumber \\
b'_1 & = & (J'_{20}+J'_{30}-J'_{10})\tanh(J_{20}+J_{30}-J_{10}).\nonumber \\
& & (cycl.)
\end{eqnarray}

The transformation for $\delta C, j''_{12}, j''_{23}, j''_{31}$ is
\begin{eqnarray}
\delta C & = & \frac{1}{4}(b''_0+b''_1+b''_2+b''_3), \nonumber \\
J''_{12}  & = & \frac{1}{4}(b''_0-b''_1-b''_2+b''_3) ~~~~(cycl.),
\end{eqnarray}
where
\begin{eqnarray}
b''_0 & = & (J'_{10}+J'_{20}+J'_{30})^2[1-\tanh^2(J_{10}+J_{20}+J_{30})]
\nonumber
\\ &&+(J''_{10}+J''_{20}+J''_{30})\tanh(J_{10}+J_{20}+J_{30}),
\nonumber \\
b''_1 & =
 &(J'_{20}+J'_{30}-J'_{10})^2[1-\tanh^2(J_{20}+J_{30}-J_{10})]
 \nonumber \\ && +(J''_{20}+J''_{30}-J''_{10})\tanh(J_{20}+J_{30}-J_{10}).
 \nonumber \\&&~~~~(cycl.)
\end{eqnarray}

The BP $\Delta$-Y transformation corresponds to inserting a new spin,
generating effective couplings $J_{10}, J_{20}, J_{30}$,
$J'_{10}, J'_{20}, J'_{30}$, and $J''_{10}, J''_{20}, J''_{30}$, such that
\begin{widetext}
\begin{eqnarray}
&& \sum_{\sigma_0}
e^{J_{10}\sigma_0\sigma_1+J_{20}\sigma_0\sigma_2+J_{30}\sigma_0\sigma_3+\delta
F}
[(J'_{10}\sigma_0\sigma_1+J'_{20}\sigma_0\sigma_2+J'_{30}\sigma_0\sigma_3+H'_{\rm ex}+\delta
U)^2 \nonumber
\\ &&+J''_{10}\sigma_0\sigma_1+J''_{20}\sigma_0\sigma_2+J''_{30}\sigma_0\sigma_3+H''_{\rm ex}+\delta C]
\nonumber \\
 &= & e^{J_{12}\sigma_1\sigma_2+J_{23}\sigma_2\sigma_3+J_{31}\sigma_3\sigma_1}
[(J'_{12}\sigma_1\sigma_2+J'_{23}\sigma_2\sigma_3+J'_{31}\sigma_3\sigma_1+H'_{\rm ex})^2 \nonumber \\
 &&+J''_{12}\sigma_1\sigma_2+J''_{23}\sigma_2\sigma_3+J''_{31}\sigma_3\sigma_1+H''_{\rm ex}], ~~~~~\forall \sigma_1,\sigma_2,\sigma_3. \nonumber \\
\label{eq:cdely}
\end{eqnarray}
\end{widetext}
Similar to the procedure in the Y-$\Delta$ transformation, we obtain
the solution for $\delta F, J_{10},..., \delta U, \cdots .$ Using the
variables $\delta f=e^{\delta F}$,
$j_i=e^{-J_{i0}},j_{ij}=e^{-J_{ij}}$, the solution for $\delta F,
J_{10},J_{20},J_{30}$ is given by
\begin{eqnarray}
j_i  & = & \sqrt{(1-t_i)/(1+t_i)},  ~~~~~~i=1,2,3 \nonumber \\
\delta f & = & z_0 /[j_1j_2j_3+1/(j_1 j_2 j_3)],
\end{eqnarray}
where
\begin{equation}
t_1=\sqrt{c_2c_3/(c_0c_1)} ~~(cycl.),
\end{equation}
with
\begin{eqnarray}
c_0 & = & z_0+z_1+z_2+z_3, \nonumber \\
c_1 & = & z_0+z_1-z_2-z_3 ~~(cycl.),
\end{eqnarray}
and
\begin{eqnarray}
z_0 & = & 1 /( j_{12}j_{23}j_{31}), \nonumber \\
z_1 & = & j_{12}j_{31} /j_{23}
 ~~(cycl.).
\end{eqnarray}

The solution for $\delta U, J'_{10},J'_{20},J'_{30}$ is given by
\begin{eqnarray}
\delta U & = & (z'_1+z'_2+z'_3-z'_0)/(c'_1+c'_2+c'_3-c'_0), \nonumber \\
J'_1 & = & \frac{1}{2}[(z'_0-z'_1)-(c'_0-c'_1)\delta U]
~~~~(cycl.),
\end{eqnarray}
with
\begin{eqnarray}
c'_0 & = & \coth(J_{10}+J_{20}+J_{30}), \nonumber \\
c'_1 & = & \coth(J_{20}+J_{30}-J_{10})~~~~ (cycl.),
\end{eqnarray}
 and
\begin{eqnarray}
z'_0 & = & c'_0(J'_{12}+J'_{23}+J'_{31}), \nonumber \\
z'_1 & = & c'_1(J'_{23}-J'_{12}-J'_{31}) ~~~~(cycl.).
\end{eqnarray}.

The solution for $\delta C, J''_{10}, J''_{20}, J''_{30}$ is given by
\begin{eqnarray}
\delta C & = & (z''_1+z''_2+z''_3-z''_0)/(c'_1+c'_2+c'_3+c'_0), \nonumber \\
J''_1 & = & \frac{1}{2}[(z''_0-z''_1)-(c'_0-c'_1)\delta C]
~~~~(cycl.),
\end{eqnarray}
where
\begin{eqnarray}
z''_0 & = & c'_0[J''_{12}+J''_{23}+J''_{31}+(c'^{-2}_0-1)(J'_{12}+J'_{23}+J'_{31})^2], \nonumber \\
z''_1 & = & c'_0[J''_{23}-J''_{12}-J''_{31}+(c'^{-2}_1-1)(J'_{23}-J'_{12}-J'_{31})^2] \nonumber \\
&&(cycl.).
\end{eqnarray}

The transformation in each step is exact. The numerical accuracy
is limited by machine's precision, which is the round-off error
$10^{-32}$ in the quadruple precision. The BP algorithm needs
about $N^3$ steps to calculate the free energy of an $N\times N$
lattice
(much faster than other numerical method). Therefore the total
error is approximately $N^{3/2}\times 10^{-32}$. This estimation
has been verified in the following way: We compared the results
obtained using double precision, in which there are 16 effective
decimal digits, and those using quadruple precision. Because the
latter results are much more accurate than the formal, we can
estimate the error in double precision results by taking the
quadruple results as the exact results. We thus found that the
error is about $N^{3/2} \times 10^{-16}$. In our calculation, the
largest size reached is $N=2000$, the round-off error is less than
$10^{-26}$.

\section{Critical internal energy and specific heat on the square
lattice with rectangular shape}

We calculate the internal energy and specific heat at critical
point $\beta_c=\frac{1}{2}\ln(1+\sqrt{2})$ with the BP algorithm.
The calculations have been carried out for aspect ratios
$\rho=M/N=1,2,4,8,16$ on an $M\times N$ rectangular lattice. For
all cases the calculation was carried out from $N=30$ to $N=2000$.
The number of data points are $120,117,117,103,92$ for
$\rho=1,2,4,8,16$, respectively.

In reference \cite{wu}, the critical internal energy density on
the square lattice with rectangle shape has been studied. The internal
energy was obtained by numerical differentiating the free energy, which was
calculated by BP algorithm for the partition function. Although
the accuracy of free energy is the machine precision, the accuracy of
internal energy can only reach $\sim 10^{-11}$. The fitting
parameters for the higher order terms for the internal energy are
not accurate enough. In the best fitting, the authors can only fit the data
up to the term $1/N^4$. More important,
the accurate form of the fitting formula can not be determined. For
example, the authors can not determine whether there exists a $\ln
N/N^{4}$ term or not. Using BP algorithm for the internal energy, the
accuracy of internal energy can reach $10^{-26}$, so we can expand the internal
energy to much higher order terms. More accurate expansion form
can be determined.

\subsection{Internal energy density}

We fit the data of critical internal energy  with the formula
given by
\begin{equation}
u=u_{\infty}+u_{\rm surf}\frac{M \ln N+N\ln M}{S} +u_{\rm corn} \frac{\ln
S }{S} +\sum_{k=1}^{\infty}\frac{B_k}{S^{k/2}}, \label{squ-inter}
\end{equation}
with $k$ from $1$ to $13$. The bulk value $u_{\infty}$ is known to
be $\sqrt{2}=1.4142135623730950488016887\cdots$ \cite{onsager}.
Our fit of $u_{\infty}$ is $1.414213562373095048801692(8)$ for
$\rho =1$, which is the worst among the five cases. The best one is
$1.414213562373095048801689(1)$ for $\rho =2$. From this one can
see the accuracy of the estimation reaches $10^{-25}$ at least.


\begin{table*}[htbp]
 \caption{ The fitted edge, corner internal energy and $B_1,B_2$ in Eq. (\ref{squ-inter}).}
\begin{tabular}{lllll}

\hline
$\rho$     & ~~~~~~~~~~$u_{\rm surf}$ & ~~~~~~~~~~$B_1$  & ~~~~~~~~~~$u_{\rm corn}$ & ~~~~~~~~~~$B_2$\\
\hline

$1$   & $-0.63661977236758134309(2)$   & $-0.1213626692058929888(4)$    & $-0.45015815807855305(3)$  & $-0.9861326825354926(3)$  \\
$2$   & $-0.636619772367581343080(3)$  & $-0.25861915103289093580(6)$   & $-0.450158158078553044(6)$  & $-1.12599588892145266(7)$ \\
$4$   & $-0.636619772367581343080(5)$  & $-0.6453979942437191475(1)$    & $-0.45015815807855305(1)$  & $-1.7143235174874347(2)$  \\
$8$   & $-0.63661977236758134308(1)$   & $-1.2664842818618753615(3)$    & $-0.45015815807855307(5)$ & $-3.2029302923645059(7)$\\
$16$  & $-0.63661977236758134309(3)$   & $-2.151538677721110286(1)$     & $-0.4501581580785531(2)$  & $ -6.492169698914722(3)$ \\
\hline
\end{tabular}
\label{fitin1}
\end{table*}

The leading correction is due to the edges. In the exact
result of Au-Yang and Fisher \cite{fisher} on the strip with two
free edges, the edges' correction is obtained as $-\frac{2}{\pi}\ln
N/N$. It is natural to conjecture that, on the rectangle with four free edges,
the edge correction is given by $-\frac{2}{\pi}(M\ln N+N\ln
M)/(MN)$ with the coefficient
\begin{equation}
u_{\rm surf}(\rho)=-\frac{2}{\pi}=-0.6366197723675813430755\dots.
\end{equation}
Our fitted $u_{\rm surf}$,  as shown in the table I, agree with this
conjecture in the accuracy $10^{-19}$.

\begin{ruledtabular}
\begin{table*}[hbtp]
\caption{The other fitted parameters of Eq. (\ref{squ-inter}) for
the critical internal energy per spin.}
\begin{tabular}{llllll}

 $\rho$   & ~~~~~~~~~~$1$ & ~~~~~~~~~~$2$ & ~~~~~~~~~~$4$ & ~~~~~~~~~~$8$ & ~~~~~~~~~~$16$\\
\hline
 $B_3 $ & $-0.29552917347574(2)$
 & $-0.098611731959506(5)$  & $~~0.84584180209494(2)$ & $ ~~3.98414088442759(8)$ & $ ~~13.5199110552941(4)$   \\
$B_4 $ & $-0.061687177276(2)$ & $-0.493056813777(8)$
 & $ -2.821652221447(4)$  &$-12.98068480072(2) $ & $ -55.3108900322(2)$ \\
$B_5 $ & $~~0.2082247526(2)$ & $~~1.100963144(1)$
 & $ ~~7.4044703357(8)$  & $~~45.243158661(7)$ & $~~265.42942328(6)$ \\
$B_6$ & $ -0.53427352(2)$ & $-2.6716642(2)$
 & $-23.0501428(1)$  & $-191.222248(2)$ & $-1557.0621850(2)$  \\
$B_7 $ & $~~1.221787(1) $ & $~~7.13298(2)$
 & $~~ 82.02634(2)$  & $~~935.7861(3)$  & $~~10631.133(5)$ \\
$B_8$ & $ -2.95720(7)$ & $-22.053(1)$
 & $-338.455(2)$  & $-5304.68(4)$ & $-83990.1(9)$  \\
$B_9 $ & $~~7.515(2) $ & $~~75.40(5)$
 & $~~1566.2(1)$  & $~~33883(4)$  & $~~7.491(1)\times 10^5$ \\
$B_{10} $ & $ -22.20(6)$ & $-294(2)$
 & $-8223(6)$  & $-2.440(3)\times 10^5$ & $-7.50(1)\times 10^6$  \\
$B_{11}$ & $~~70(1) $ & $~~1.24(5)\times 10^3$
 & $~~4.66(2)\times 10^4$  & $~~1.89(1)\times 10^6$  & $~~8.12(8)\times 10^7$ \\
$B_{12} $ & $ -220(10)$ & $-5.1(6)\times 10^3$
 & $-2.58(4)\times 10^5$  & $-1.44(4)\times^7$ & $-8.5(3)\times 10^8$  \\
$B_{13} $ & $~~410(40) $ & $~~1.3(4)\times 10^4$
 & $~~9.4(4)\times 10^5$  & $~~7.3(4)\times 10^7$  & $~~6.0(5)\times 10^9$ \\
 \hline
\end{tabular}

\label{fitien}
\end{table*}
\end{ruledtabular}

The term $B_1/S^{1/2}$ in fact scales as $1/N$. In the infinitely
long strip limit, the coefficient of $1/N$ is known as
$\frac{2}{\pi}(\frac{7}{2}\ln 2+\gamma-\frac{\pi}{4}-\ln
\pi)\approx 0.683158$ \cite{fisher}. As we can see in Tab.
\ref{fitin1}, $B_1/\sqrt{\rho}$ indeed approaches this limit as
$\rho$ increases.

Following the convention in the critical free energy, we write the
coefficient of $(\ln S)/S$ as $u_{\rm corn}$. From the fitted
$u_{\rm corn}$ in table I, we conjecture that it equals to
\begin{equation}
u_{\rm corn}(\rho)=-\frac{\sqrt{2}}{\pi}=-0.4501581580785530347\cdots,
\label{eq:u-corn}
\end{equation}
which is independent of aspect ratio $\rho$, hence geometry
independent. The accuracy of this result reaches $10^{-16}$, much
higher than that in the previous work \cite{wu}.

The other parameters $B_2, B_3, \cdots$ are fitted and listed in
Tab. \ref{fitin1} and II. We have tried other forms of formula to fit the
critical internal energy. The logarithmic correction in higher
order terms such as $\ln S/S^{3/2}, \ln S/S^2, \cdots$ are
excluded clearly since their coefficients are extremely small if
these terms are taken into account. Moreover the standard
deviations of the fits with these terms are much larger than those
without them.

\subsection{Critical specific heat density}

The data of the critical specific heat are fitted according to the
formula
\begin{eqnarray}
c&=&A^{\rm squ}_0 \ln N+c_0+c_{\rm surf}\frac{M \ln N+N\ln M}{S} +c_{\rm corn}
\frac{\ln S }{S} \nonumber \\
&&+\sum_{k=1}^{\infty}\frac{D_k}{S^{k/2}}, \label{squ-spe}
\end{eqnarray}
with $k$ from $1$ to $13$. The leading term $A^{\rm squ}_0\ln N$ is known
from Onsager's exact result \cite{onsager}, which reads
\begin{equation}
A^{\rm squ}_0=\frac{2}{\pi}[\ln(1+\sqrt{2})]^2= 0.4945385895323191178650031
\cdots.
\end{equation}
Among the five aspect ratios, the worst fit gives $A^{\rm squ}_0\approx
0.49453858953231911786503(9)$, which agrees with the exact result
in the accuracy $10^{-22}$. Other fitted parameters are listed
in Tab.~III and IV.

\begin{table*}[htbp]
 \caption{ The fitted edge, corner specific heat and $c_0$ in Eq. (\ref{squ-spe}).  }
\begin{tabular}{lllll}

\hline
$\rho$    & ~~~~~~~~~~$c_0$ & ~~~~~~~~~~~$c_{\rm surf}$ & ~~~~~~~~~~$D_1$  & ~~~~~~~~~~$c_{\rm corn}$ \\
\hline

$1$   & $-0.5707862077066987030073(9)$  & $0.5245373853251001249(3)$    & $-0.349418614766849034(2)$  & $0.3709039421492394(2)$   \\
$2$   & $-0.4427629983689372851264(1)$  & $0.52453738532510012479(4)$   & $-0.6263517310106051186(6)$ & $0.37090394214923938(3)$ \\
$4$   & $-0.3776612153104486912262(1)$  & $0.52453738532510012478(6)$   & $-0.445730673715527357(1)$  & $0.37090394214923940(9)$ \\
$8$   & $-0.3451074842559496760002(3)$  & $0.5245373853251001248(1)$    & $-0.190599347476723077(3)$  & $0.3709039421492395(3)$\\
$16$  & $-0.3288306187123204989915(1)$  & $0.52453738532510012481(4)$   & $~~0.132303803774576619(1)$   & $0.3709039421492396(2)$  \\
 \hline
\end{tabular}
\label{squ-spe-ta}
\end{table*}

The constant $c_0$ increases with the aspect ratio. For the
strip case it is known that $-c_0(\rho=\infty)=(\frac{7}{2}\ln
2+\gamma-14\xi(3)/\pi^2-\frac{\pi}{4}-\ln \pi)\approx 0.3125538$
\cite{fisher}. $c_0$ approaches this limit as $\rho \to \infty$
obviously, see Tab. III. However we have not obtained an
analytical expression for the dependence of $c_0$ on $\rho$.

\begin{ruledtabular}
\begin{table*}[hbtp]
\caption{The fitted parameters of Eq. (\ref{squ-spe}) for the
critical specific heat per spin. }
\begin{tabular}{llllll}
 $\rho$   & ~~~~~~~~~~$1$ & ~~~~~~~~~~$2$ & ~~~~~~~~~~$4$ & ~~~~~~~~~~$8$ & ~~~~~~~~~~$16$\\
\hline

$D_2 $ & $~~1.126498783937362(2)$
 & $~~1.2893849732698274(4)$  & $~~1.904772120606373(1)$ & $ ~~3.393198964111045(4)$ & $ ~~6.627143703373272(2)$   \\
$D_3 $ & $~~0.15426462287065(7)$ & $-0.09877525669444(2)$
 & $ -1.15535182820573(7)$  &$-4.5040900875957(3) $ & $ -14.4878278694865(2)$ \\
$D_4 $ & $~~0.072511922527(6)$ & $~~0.507330707955(3)$
 & $ ~~2.85076699663(1)$  &$~~13.05859574667(8) $ & $ ~~55.54544991595(6)$ \\
$D_5 $ & $-0.2918984491(6)$ &
$-1.1746661841(3)$ & $ -7.601311331(2)$  & $-45.75496161(2)$ & $-266.6226024501(6)$ \\
$D_6  $ & $ ~~0.64864572(5)$ & $~~2.77747936(4)$
 & $~~23.0088946(3)$  & $~~187.261430(4)$ & $~~1510.853814(6)$  \\
$D_7  $ & $-1.460056(3) $ & $-7.593353(3)$
 & $-80.89031(4)$  & $-888.1909(7)$  & $-9896.112(1)$ \\
$D_8 $ & $ ~~3.4124(1)$ & $~~23.1018(2)$
 & $~~321.261(3)$  & $~~4762.20(8)$ & $~~73171.3(2)$  \\
$D_9 $ & $-9.608(4) $ & $-82.83(1)$
 & $-1480.6(3)$  & $-29095(8)$  & $-6.0847(3)\times 10^5$ \\
$D_{10} $ & $ ~~30.7(1)$ & $~~337.3(3)$
 & $~~7.76(1)\times 10^3$  & $~~1.998(5)\times 10^5$ & $~~5.617(2)\times 10^6$  \\
$D_{11} $ & $-113(2) $ & $-1591(7)$
 & $-4.68(4)\times 10^4$  & $-1.55(2)\times 10^6$  & $-5.72(1)\times 10^7$ \\
$D_{12} $ & $ ~~380(20)$ & $~~7.1(1)\times 10^3$
 & $~~2.76(7)\times 10^5$  & $~~1.19(6)\times 10^7$ & $~~5.72(5)\times 10^8$  \\
$D_{13} $ & $-900(70) $ & $-2.30(6)\times 10^5$
 & $-1.18(6)\times 10^6$  & $-6.7(7)\times 10^7$  & $-4.16(9)\times 10^9$ \\

\hline
\end{tabular}

\label{fitspe2}
\end{table*}
\end{ruledtabular}

The term $(M\ln N+N\ln M)/S$ is the next order correction. From
the fit of $c_{\rm corn}$, listed in table \ref{squ-spe-ta}, one can
see that it is independent of $\rho$.  We conjecture that its
exact value is given by
\begin{equation}
c_{\rm surf}=\frac{3\sqrt{2}}{4}A^{\rm squ}_0=0.5245373853251001248(3).
\end{equation}
In the previous work it is given by $c_{\rm surf}=0.524529(3)$
\cite{wu}. Note that this term is absent in the torus case
\cite{fisher1969} and not mentioned in the infinitely long strip case
\cite{fisher}, but exists in the cylinder case with Brascamp-Kunz
boundary conditions \cite{Janke,izmailian2002b}.

The corner term $c_{\rm corn}$ seems also independent of aspect ratio
$\rho$ and its exact value is conjectured as
\begin{equation}
c_{\rm corn}=\frac{3}{4}A^{\rm squ}_0=0.370903942149239338\cdots .
\end{equation}

The other parameters $D_1, D_2, \cdots$ are fitted and listed in
Tab. \ref{fitspe2}. We have tried other forms of formula to fit the specific
heat. The logarithmic correction in higher order terms such as
$\ln S/S^{3/2}, \ln S/S^2, \cdots$ are excluded clearly since
their coefficients are extremely small if these terms are taken
into account. Moreover the standard deviations of the fits with
these terms are much larger than those without them.

\section{Critical internal energy and specific heat on the triangular lattice with free boundary}

For the triangular lattice, we have studied five shapes: triangle,
rhombus, trapezoid, hexagon and rectangle, as shown in Fig. 2.
Using the BP algorithm for the internal energy and specific heat,
we obtain the critical energy density and specific heat for the
five shapes at the exact critical point
$\beta_c=\frac{1}{4}\ln(3)=0.274653072167\cdots$. The linear size
$N$ of a finite lattice is defined as the length of edges in the
triangle, rhombus and hexagon cases, of which the length of edges
are equal. For the trapezoid shape,  the lengths of the three
shorter edges are required to be equal and $N$ is the length of
the shorter edges. For the rectangle, $N$ is defined as the length
of the bottom edge, and the number of layers is also required to
be $N$. However, the actual geometrical vertical length is
$N\sqrt{3}/2$. According to the finite-size scaling
\cite{privman}, the system size should be the actual geometrical
length, not the number of layers. Therefore the aspect ratio of
the rectangle, we consider here, is $\sqrt{3}/2$ rather than $1$.

The area $S$, i.e. the number of spins on the lattice, is given by
$S=N(N+1)/2$, $N^2$, $N(3N-1)/2$, $3N^2-3N+1$, $N^2-(N-1)/2$ for
the triangle, rhombus, trapezoid, hexagon and rectangle shaped
system, respectively.

\subsection{Critical internal energy density}

We fit the data of critical internal energy with the formula given
by
\begin{eqnarray}
u&=&u_{\infty}+u_{\rm surf}\frac{p(N) \ln N}{S}+\frac{u_1}{S^{1/2}} +
\frac{u_{\rm corn}\ln S+u_2 }{S} \nonumber \\
&&+\sum_{k=3}^{\infty}\frac{u_k+u_{lk}\ln N}{S^{k/2}},
\label{eq:tri-inter}
\end{eqnarray}
where $p(N)$ is the perimeter, which equals to $3N, 4N, 5N-1, 6N,
(2+\sqrt{3})N$ for the triangle, rhombus, trapezoid, hexagon and
rectangle, respectively. This expansion is different from that for
the triangular lattice with periodic boundary conditions
\cite{salas}, in which there's no surface, corner terms,
logarithmic corrections, and only even $k$ presents. In addition
there are logarithmic correction in higher order terms such as
$\ln S/S^2, \ln S/S^{5/2}, \cdots$, for the rhombus, trapezoid,
and hexagon shapes, while such terms are absent for the triangle
and rectangle shape. In the previous work \cite{wu2}, we have found
the logarithmic corrections in higher order terms in the critical
free energy for the rhombus, trapezoid and hexagon shapes.

\begin{figure}
\includegraphics[width=0.5\textwidth]{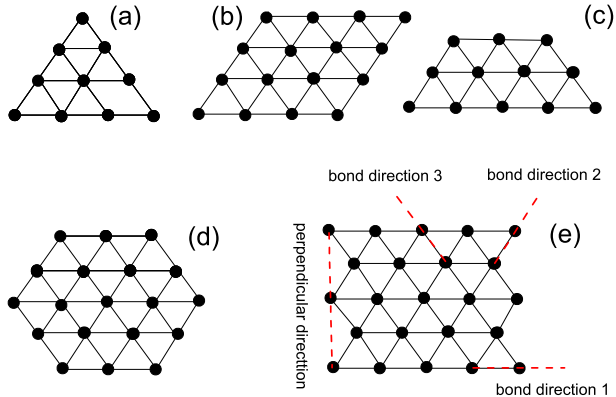}
\caption{ (a) The triangle-shaped triangular lattice with $N=4$.
(b) The rhombus-shaped lattice with $N=4$.    (c) The
trapezoid-shaped lattice with $N=3$. (d) The hexagon-shaped
lattice with $N=3$. (e) The rectangle-shaped lattice with $N=5$.
Three bond directions and the perpendicular direction are shown (see text).}
\label{shape}
\end{figure}

\begin{table*}[htbp]
 \caption{ The fitted edge, corner internal energy and $u_k$ for triangle and rectangle shape in Eq. (\ref{eq:tri-inter}).
 For triangle and rectangle shape, there is no other logarithmic corrections than surface and corner terms. }
\begin{tabular}{lll}

\hline
            & ~~~~~~~~~~ triangle               &  ~~~~~~~~~~rectangle                   \\
\hline
$u_{\rm surf}$   & $-0.551328895421792049511325(9)$   & $-0.5513288954217920495113(1)$       \\
$u_1 $       & $~~~0.1358128850971729047321(2)$      & $-1.124623054881823516433(4))$         \\
$u_{\rm corn}$   & $-1.65398668626537614853(3)$       & $~~~0.1477281322922120845(6)$               \\
$u_2 $       & $-1.4200975898148873036(2)$        & $-1.236099923077146896(3)$         \\
$u_3 $       & $-0.702623731123014999(6)$         & $~~~0.3520087573021713(2)$         \\
$u_4 $       & $-0.0272494452610576(5)$           & $-0.04481963510134(2)$         \\
$u_5 $       & $~~~0.01706870885164(4)$              & $~~~0.141108181792(2)$         \\
$u_6 $       & $-0.009549023082(3)$               & $~~~0.0055256275(2)$         \\
$u_7 $       & $~~~0.0102157015(1)$                  & $~~~0.10191756(1)$         \\
$u_8 $       & $-0.009092781(5)$                  & $-0.2825362(7)$         \\
$u_9 $       & $~~~0.0024505(1)$                     & $~~~0.40416(3)$         \\
$u_{10} $    & $~~~0.006205(2)$                      & $-0.7968(7)$         \\
$u_{11} $    & $-0.01636(3)$                      & $~~~0.04(1)$         \\
$u_{12} $    & $~~~0.0274(2)$                        & $~~~1.4(1)$         \\
$u_{13} $    & $-0.0232(6)$                       & $-7.9(6)$         \\
\hline
\end{tabular}
\label{tab:tri-inter1}
\end{table*}

For triangle and rectangle-shaped lattices, we fit the data with $k$ from
$1$ to $13$. We find that there is no logarithmic correction in
higher orders, i.e., $u_{lk}=0$.
For rhombus, trapezoid and
hexagon shape, we fit the data with $k$ from $1$ to $9$. We find
that there are logarithmic corrections in higher orders. The
higher order logarithmic corrections begins from $k=4$ as listed in
Tab. \ref{tab:tri-inter2}.

The bulk value $U_{\infty}$ is known to be $2$ \cite{newell}. Our
fit of $U_{\infty}$ is $2.0+(3\pm 9)\times 10^{-23}$ for the
hexagonal shape, which is the worst among five cases. The best one
is $2.0 \pm 10^{-25}$ for the triangle shape. From this one can see
the accuracy of the BP algorithm reaches $10^{-25}$ at least.

The leading correction is due to the edges (or surfaces), which is
unknown to our knowledge. We denote the surface internal energy
per unit length of the edge along one bond direction by
$u^{\parallel}_{\rm surf}$, and that perpendicular to that direction
by $u^{\perp}_{\rm surf}$. For the triangular, rhomboid, trapezoid,
and hexagonal shapes (see Table \ref{tab:tri-inter1} and
\ref{tab:tri-inter2}) , we have $u_{\rm surf}=u_{\rm surf}^{\parallel}$.
For the rectangular shape we have $u_{\rm surf}=(2
u_{\rm surf}^{\parallel}+\sqrt{3}u^{\perp}_{\rm surf})/(2+\sqrt{3})$, and
\begin{eqnarray}
u_{\rm surf}^{\parallel} & = &-0.551328895421792049511325(9),
\nonumber \\
    u_{\rm surf}^{\perp} & = & -0.5513288954217920495113(1).
\end{eqnarray}
This is an interesting result because it means that the surface
internal energy per unit length is symmetric for an edge along or
perpendicular to an arbitrary bond direction. We conjecture that
the exact value of the surface internal energy  is given by
\begin{equation}
u_{\rm surf}^{\parallel}=u_{\rm surf}^{\perp}
=-\frac{\sqrt{3}}{\pi}=-0.551328895421792049511326 \cdots
\end{equation}

\begin{table*}[htbp]
 \caption{ The fitted edge, corner internal energy and $u_k$ for rhombus, trapezoid and hexagon shape in Eq. (\ref{eq:tri-inter}).  }
\begin{tabular}{llll}

\hline
            & ~~~~~~~~~~ rhombus               & ~~~~~~~~~~~trapezoid                & ~~~~~~~~~~~hexagon \\
\hline

$u_{\rm surf}$  & $-0.55132889542179204946(7)$   & $-0.551328895421792049501(7)$   & $-0.5513288954217920497(5)$   \\
$u_1$       & $-0.728607166200085890(3)$     & $-1.4553104976350188627(3)$      & $-3.44983946582806485(3)$     \\
$u_{\rm corn}$  & $-0.7351051938957218(9)$       & $-0.7351051938957225(1)$         & $~~1.102657790843582(8)$     \\
$u_2$       & $-1.093360446288389(6)$        & $-0.5538076013331216(8)$         & $~~ 1.71029326576843(5)$     \\
$u_3$       & $-0.219739754536(1)$           & $~~0.0257228180889(2)$           & $-0.37845053923(2)$      \\
$u_4$       & $~~0.067693864(4)$             & $-0.0117835146(8)$               & $~~0.34795211(8)$     \\
$u_5$       & $-0.016510(1)$                 & $-0.0171482(3)$                  & $-0.28066(3)$     \\
$u_6$       & $~~0.0046(1)$                  & $~~0.09999(2)$                   & $~~0.271(4)$     \\
$u_7$       & $-0.113(3)$                    & $-0.2641(8)$                     & $-2.6(2)$     \\
$u_8$       & $~~0.14(1)$                    & $~~0.446(5)$                     & $-4(2)$     \\
$u_9$       & $-0.10(1)$                     & $-0.549(4)$                      & $-3.8(9)$     \\
\hline
$u_{l4}$       & $~~0.0335649470(7)$              & $~~0.0420829435(1)$             & $~~0.09678630(1)$     \\
$u_{l5}$       & $-0.0223773(2)$                & $-0.04922954(5)$              & $ -0.055879(7)$     \\
$u_{l6}$       & $~~0.01797(3)$                   & $~~0.059675(7)$                 & $~~0.065(1)$     \\
$u_{l7}$       & $~~0.0104(9)$                    & $-0.0450(3)$                  & $~~0.55(6)$     \\
$u_{l8}$       & $~~0.012(9)$                     & $~~0.025(3)$                    & $~~5(1)$     \\
$u_{l9}$       & $~~0.06(1)$                      & $~~0.037(7)$                    & $~~15(3)$     \\
\hline
\end{tabular}
\label{tab:tri-inter2}
\end{table*}

Following the convention for the critical free energy, we write
the coefficient of $(\ln N)/S$ as $u_{\rm corn}$. We denote the corner
correction by $u_{\rm corn}^{(\gamma)}$, where $\gamma$ is the angle
of the corner. Under the assumption that $u_{\rm corn}$ is the sum of
the corner's contributions, we have
\begin{eqnarray}
u_{\rm corn}=\left\{\begin{array}{lcl} 3 u_{\rm corn}^{(\pi/3)}
\hspace{2.2cm} \mbox{for triangle}
\nonumber\\
2\left(u_{\rm corn}^{(\pi/3)}+u_{\rm corn}^{(2\pi/3)}\right)  \hspace{0.2cm} \mbox{for rhombus and trapezoid} \nonumber\\
6 u_{\rm corn}^{(2\pi/3)} \hspace{2.1cm} \mbox{for hexagon}
\nonumber\\
 4 u_{\rm corn}^{(\pi/2)} \hspace{2.2cm} \mbox{for rectangle} \nonumber
\end{array}
\right.
\end{eqnarray}

From the Tab. \ref{tab:tri-inter1} and \ref{tab:tri-inter2}, we
obtain the corner term for the three angles
\begin{eqnarray}
u_{\rm corn}^{(\pi/3)}& = & -0.55132889542179204951(1)\approx -\frac{\sqrt{3}}{\pi} ,\nonumber \\
u_{\rm corn}^{(\pi/2)}& = & 0.036932033073053021125(2),\nonumber \\
u_{\rm corn}^{(2\pi/3)}& = & 0.1837762984739298(3) \approx
\frac{1}{\sqrt{3}\pi}.
\end{eqnarray}
From this result, we can conjecture that the exact result of
$u_{\rm corn}^{(\pi/3)}$ is $-\sqrt{3}/\pi$, $u_{\rm corn}^{(2\pi/3)}$ is
$1/(\sqrt{3}\pi)$

There is no logarithmic corrections in higher order terms for the
triangular and rectangular shapes. In contrast there are such
terms for the rhombus, trapezoid and hexagon shapes. With the help of much
more accurate data, we confirm the finding in the previous work \cite{wu2}
that, for in critical free energy density, there is no logarithmic correction
terms of order higher
than $1/S$ for the triangular and rectangular shape, and there are
such terms for the rhombus, trapezoid and hexagonal shape.

\subsection{Critical specific heat density}

The data of the critical specific heat are fitted using the
following formula
\begin{eqnarray}
c&=&A^{\rm tri}_0\ln N+c_0+c_{\rm surf} \frac{p(N)\ln N}{S}+\frac{c_1}{S^{1/2}}
\nonumber
\\ &&+\frac{c_{\rm corn}\ln N+c_2}{S} +\sum_{k=3}\frac{c_k+c_{lk}\ln
N}{S^{k/2}}, \label{eq:tri-spe}
\end{eqnarray}
where $p(N)$ is the perimeter. Compared with the expansion for the
triangular lattice with periodic boundary conditions \cite{salas},
there are additional logarithmic surface term and corner term.

\begin{table*}[htbp]
 \caption{ The fitted edge, corner internal specific heat and $c_k$ for triangle and rectangle shape in Eq. (\ref{eq:tri-spe}).  }
\begin{tabular}{lll}

\hline
            & ~~~~~~~~~~ triangle               &  ~~~~~~~~~~rectangle                   \\
\hline
$A^{\rm tri}_0     $   & $~~0.499069378046460449833662(8)$    & $~~0.4990693780464604498336723(9)$       \\
$c_0     $   & $-0.80424240621924908720053(9)$      & $-0.573388979085776611481176(9)$       \\
$c_{\rm surf}$   & $~~0.166356459348820149932(8)$       & $~~0.138096991677687094105(1)$       \\
$c_1 $       & $-0.1878730393246832268(2)$          & $-0.08534785228081183795(4))$         \\
$c_{\rm corn}$   & $~~0.74860406706969065(1)$           & $-0.150040848081735954(3)$               \\
$c_2 $       & $~~0.64050897872895593(7)$           & $~~0.68035027281155164(2)$         \\
$c_3 $       & $~~0.341267433764613(2)$             & $-0.0382477471847520(7)$         \\
$c_4 $       & $~~0.0164170250248(1)$               & $~~0.06815513150821(7)$         \\
$c_5 $       & $-0.008335950854(7)$                 & $-0.027408848314(6)$         \\
$c_6 $       & $~~0.0048860618(4)$                  & $~~0.0752274355(5)$         \\
$c_7 $       & $-0.00632229(1)$                     & $-0.16560769(3)$         \\
$c_8 $       & $~~0.0050319(4)$                     & $~~0.207074(1)$         \\
$c_9 $       & $-0.002319(5)$                       & $-0.28531(5)$         \\
$c_{10} $    & $-0.032(2)$                          & $-0.112(1)$         \\
$c_{11} $    & $~~0.43(4)$                          & $~~0.85(2)$         \\
$c_{12} $    & $-2.8(3)$                            & $-3.6(2)$         \\
$c_{13} $    & $~~8(1)$                             & $~~3.3(9)$         \\
\hline
\end{tabular}
\label{tab:tri-spe1}
\end{table*}

For triangle and rectangle shape, we fit the data with $k$ from
$1$ to $13$. We find that there is no logarithmic correction in
higher order terms, i.e. $c_{lk}=0$. Therefore there is no
$c_{lk}$ in Tab \ref{tab:tri-spe1}. For rhombus, trapezoid and
hexagon shape, we fit the data with $k$ from $1$ to $9$. We find
that there are logarithmic correction in higher order terms. The
higher order logarithmic corrections begins from $k=3$ as shown in
Tab. \ref{tab:tri-spe2}.

The leading term $A^{\rm tri}_0\ln N$ is known from the exact result
\cite{salas}, which reads
\begin{equation}
A^{\rm tri}_0=\frac{3\sqrt{3}}{4\pi}(\ln3)^2=
0.4990693780464604498336724\cdots.
\end{equation}
Our fits, listed in Tab. \ref{tab:tri-spe1} and \ref{tab:tri-spe2}, agree
with the exact result at least in accuracy $10^{-21}$. The other
fitted parameters are listed in Tab. \ref{tab:tri-spe1} and
\ref{tab:tri-spe2} as well.

\begin{table*}[htbp]
 \caption{ The fitted edge, corner internal specific heat and $c_k$ for rhombus, trapezoid and hexagon shape in Eq. (\ref{eq:tri-spe}).  }
\begin{tabular}{llll}

\hline
            & ~~~~~~~~~~ rhombus               & ~~~~~~~~~~~trapezoid           & ~~~~~~~~~~~hexagon \\
\hline
$A^{\rm tri}_0$       & $~~0.499069378046460449832(2)$   & $~~0.4990693780464604498337(1)$    & $~~0.499069378046460449831(1)$     \\
$c_0$       & $-0.60510340124288164836(2)$     & $-0.519950222059951425774(2)$      & $-0.26733979383373518196(1)$     \\
$c_{\rm surf}$  & $~~0.166356459348820142(8)$      & $~~0.1663564593488201503(8)$       & $~~0.166356459348820129(9)$   \\
$c_1$       & $-0.146850680614596(3)$          & $-0.10404207253293299(4)$          & $-0.2755298974420975(5)$     \\
$c_{\rm corn}$  & $~~0.3881650718138(1)$           & $~~0.38816507181392(1)$            & $-0.3327129186981(1)$     \\
$c_2$       & $-1.093360446288389(6)$          & $~~0.6373742819503(8)$             & $~~0.543662728665(1)$     \\
$c_3$       & $~~0.1409052856(8)$              & $-0.023617749(8)$                  & $-0.0373484(1)$      \\
$c_4$       & $~~0.00563(3)$                   & $-0.0117835146(8)$                 & $~~0.34795211(8)$     \\
$c_5$       & $-0.016510()$                    & $~~0.027341(1)$                    & $~~0.03672(2)$     \\
$c_6$       & $~~-0.045(2)$                    & $-0.0232(1)$                       & $~~0.181(3)$     \\
$c_7$       & $-0.00(3)$                       & $~~0.084(3)$                       & $~~2.24(9)$     \\
$c_8$       & $~~0.2(1)$                       & $~~0.40(3)$                        & $~~16(1)$     \\
$c_9$       & $~~0.2(1)$                       & $~~0.95(5)$                        & $~~39(3)$     \\
\hline
$c_{l3}$       & $~~0.0269426979(1)$              & $~~0.02995964044(2)$          & $~~0.0787927005(2)$     \\
$c_{l4}$       & $-0.01404491(5)$                 & $~~0.00186976(4)$             & $-0.0398024(6)$     \\
$c_{l5}$       & $~~0.011518(7)$                  & $~~0.011667(7)$               & $~~0.0418(1)$     \\
$c_{l6}$       & $~~0.0009(4)$                    & $-0.0626(4)$                  & $-0.859(9)$     \\
$c_{l7}$       & $~~0.04(1)$                      & $~~0.021(8)$                  & $-3.1(3)$     \\
$c_{l8}$       & $~~0.15(8)$                      & $-0.82(4)$                    & $-21(2)$     \\
$c_{l9}$       & $~~0.1(1)$                       & $~~0.57(4)$                   & $~~11(1)$     \\
\hline
\end{tabular}
\label{tab:tri-spe2}
\end{table*}

The leading correction $p(N)\ln N/S $ is caused by the edges. We
denote the surface specific heat per unit length of the edge along
one bond direction by $c^{\parallel}_{\rm surf}$, and that
perpendicular to the direction by $c^{\perp}_{\rm surf}$. For the
triangle, rhombus, trapezoid, and hexagon shape, we have
$c_{\rm surf}=c_{\rm surf}^{\parallel}$;  for the rectangle, we set
$c_{\rm surf}=(2
c_{\rm surf}^{\parallel}+\sqrt{3}c^{\perp}_{\rm surf})/(2+\sqrt{3})$, and
find
\begin{eqnarray}
c_{\rm surf}^{\parallel} & = & 0.166356459348820149932(8)\approx
\frac{A^{\rm tri}_0}{3}, \nonumber
\\
    c_{\rm surf}^{\perp} & = &0.105465769143518701148(8).
\end{eqnarray}
Note that this term is absent in the torus case \cite{fisher1969}
and not mentioned in the long strip case \cite{fisher}, but exists
in the cylinder case with Brascamp-Kunz boundary conditions
\cite{Janke,izmailian2002b}. In addition, we
conjecture that
$c_{\rm surf}^{\parallel}=A^{\rm tri}_0/3$.

Following the convention for the critical free energy, we write
the coefficient of $\ln N/S$ as $c_{\rm corn}$. We denote the corner
correction by $c_{\rm corn}^{(\gamma)}$ where $\gamma$ is the angle of
the corner. Again, under the assumption that the total correction
is the sum of all corners,  we have
\begin{eqnarray}
c_{\rm corn}=\left\{\begin{array}{lcl} 3 c_{\rm corn}^{(\pi/3)}
\hspace{2.2cm} \mbox{for triangle}
\nonumber\\
2\left(c_{\rm corn}^{(\pi/3)}+c_{\rm corn}^{(2\pi/3)}\right)~~ \mbox{for rhombus and trapezoid} \nonumber\\
6 c_{\rm corn}^{(2\pi/3)} \hspace{2.1cm} \mbox{for hexagon}
\nonumber\\
 4 c_{\rm corn}^{(\pi/2)}\hspace{2.2cm} \mbox{for rectangle} \nonumber
\end{array}
\right.
\end{eqnarray}
From Tab. \ref{tab:tri-spe1} and \ref{tab:tri-spe2}, we obtain the corner contribution of the
three angles,
\begin{eqnarray}
c_{\rm corn}^{(\pi/3)}& = & 0.249534689023230217(3)\approx \frac{A^{\rm tri}_0}{2},\nonumber \\
c_{\rm corn}^{(\pi/2)}& = &-0.0375102120204339885(8),\nonumber \\
c_{\rm corn}^{(2\pi/3)}& = &-0.05545215311627(5)\approx \frac{A^{\rm tri}_0}{9}.
 \label{eq:c-corn}
\end{eqnarray}
We conjecture that $c_{\rm corn}^{(\pi/3)}= A^{\rm tri}_0/2$ and
$c_{\rm corn}^{(2\pi/3)}=A^{\rm tri}_0/9$.

In the previous section on the square lattice, we obtained the
corner term $c_{\rm corn}=0.3709039421492394(2)$ for the rectangular
shape with various aspect ratios. It is different from the
present result $-0.150040848081735954(3)$ for the rectangle-shaped
triangular lattice. This indicates that the corner term of the
specific heat depends on the microscopic structure of the lattice,
thus is not universal.

\section{Summary and discussion }

We have developed the BP algorithm for the specific heat and made
the BP algorithm for the internal energy completed. With these
algorithm, we have studied the surface and corner quantities via a
numerically exact approach on 2D lattices with free boundaries,
which permits us to extract very precisely many terms in their
asymptotic expansions, and sometimes even allows conjectures of
their exact values. This work is a continuation of earlier works
\cite{wu, wu2} by the three of the present authors, but here
addressing specifically the internal energy and the specific heat.
There are two remarkable progresses comparing with previous work
\cite{wu,wu2}:

I. Some exact edge and corner terms are conjectured from the
accurate numerical fits. For the rectangular shape on the
square lattice, the exact edge term and corner term of internal
energy and specific heat $u_{\rm surf},u_{\rm corner},c_{\rm surf},c_{\rm corn}$
are conjectured. From the results for various shapes on the
triangular lattice, we conjectured the exact edge and corner terms
$u_{\rm surf}^{\parallel}$, $u_{\rm surf}^{\perp}$,$u_{\rm corn}^{(\pi/3)}$,
$u_{\rm corn}^{(2\pi/3)}$,
$c_{\rm surf}^{\parallel}$,$c_{\rm corn}^{(\pi/3)}$,
$c_{\rm corn}^{(2\pi/3)}$. These conjectured exact results imply that
there should exist closed form analytical solutions for these cases with
free boundaries.

II. The accurate forms of finite-size scaling for the internal
energy and specific heat are determined. For the rectangular shape
on the square and triangular lattice, there is no logarithmic
correction terms of order higher than $1/S$. For the triangle
shape on the triangular lattice, there is also no logarithmic
correction terms higher that $1/S$. For the rhombus, trapezoid and
hexagonal shape on the triangular lattice, there exist logarithmic
correction terms of order higher than $1/S$ for the internal
energy, and logarithmic correction terms of all orders for the
specific heat. This property seems harmonic. The origin of these
higher order logarithmic corrections are an interesting topic for
the RG and CFT.

With Vernier and  Jacobsen's  analytical solution \cite{jacobsen}, one can
study the internal energy and specific heat. However this calculation
has not been carried out. It should be interesting to compare the results
obtained by their solution and our numerical results with the BP algorithm.

{\it acknowledgment}
This work is supported by the National Science Foundation of China (NSFC)
under Grant No. 11175018. N.I. is also supported by FP7 EU IRSES Project No. 295302 (Statistical Physics in
Diverse Realizations) and by a Marie Curie International Incoming Fellowship within
the 7th European Community Framework Programme.

\end{document}